\documentclass[sigplan,screen]{acmart}
\usepackage[utf8]{inputenc}
\usepackage{float}
\usepackage{graphicx}
\usepackage{caption}
\usepackage{import}
\usepackage{tikz}

\usepackage{subcaption}

\usepackage{mathtools}
\usepackage{wasysym}

\usepackage{placeins} 

\usepackage[linesnumbered,ruled,vlined]{algorithm2e}
\SetKwInput{KwInput}{Input}              
\SetKwInput{KwOutput}{Output}
\usepackage{xurl}
\usepackage{multirow}
\usepackage{tablefootnote}

\usepackage{listings}

\definecolor{codegreen}{rgb}{0,0.7,0}
\definecolor{codegray}{rgb}{0.5,0.5,0.5}
\definecolor{codered}{rgb}{0.9,0.3,0.3}
\definecolor{backcolour}{rgb}{0.92,0.92,0.92}

\definecolor{codegreen}{rgb}{0,0.6,0}
\definecolor{codegray}{rgb}{0.5,0.5,0.5}
\definecolor{codepurple}{rgb}{0.58,0,0.82}
\definecolor{backcolour}{rgb}{0.95,0.95,0.92}
\definecolor{codered}{rgb}{0.6,0,0}
\definecolor{codeblue}{rgb}{0,0,0.8}
\definecolor{codewhite}{rgb}{1,1,1}

\lstdefinestyle{mystyle}{
    frame=tb,
    backgroundcolor=\color{codewhite}, 
    commentstyle=\color{codegreen},
    keywordstyle=\color{blue},
    numberstyle=\tiny\color{codegray},
    stringstyle=\color{codered},
    basicstyle=\ttfamily\scriptsize,
    breakatwhitespace=false,         
    breaklines=true,                 
    captionpos=b,                    
    keepspaces=true,                 
    numbers=left,                    
    numbersep=5pt,                  
    showspaces=false,                
    showstringspaces=false,
    showtabs=false,                  
    tabsize=1,    
    belowskip=0pt, 
    xleftmargin=-2pt,  
    xrightmargin=-2pt,
    morekeywords={.data,.global,.quad,.space,.macro,.endm,mfence,rdtsc,shl,add,sub,pop,push,ret,lea,.text,.align,.rept,.endr}
}

\copyrightyear{2025}
\acmYear{2025}
\setcopyright{cc}
\setcctype{by}
\acmConference[ASPLOS '25]{Proceedings of the 30th ACM International Conference on Architectural Support for Programming Languages and Operating Systems, Volume 2}{March 30-April 3, 2025}{Rotterdam, Netherlands}
\acmBooktitle{Proceedings of the 30th ACM International Conference on Architectural Support for Programming Languages and Operating Systems, Volume 2 (ASPLOS '25), March 30-April 3, 2025, Rotterdam, Netherlands}\acmDOI{10.1145/3676641.3716274}
\acmISBN{979-8-4007-1079-7/25/03}



\begin{document}
\title{SMaCk: Efficient Instruction Cache Attacks via Self-Modifying Code Conflicts}

\author{Seonghun Son}
\orcid{0009-0009-9869-0970}
\affiliation{%
  \institution{Iowa State University}
  \country{USA}
}
\email{seonghun@iastate.edu}

\author{Daniel Moghimi}
\orcid{0000-0002-3123-5916}
\affiliation{%
  \institution{Google}
  \country{USA}
}
\email{danielmm@google.com}

\author{Berk Gulmezoglu}
\orcid{0000-0001-6268-6325}
\affiliation{%
  \institution{Iowa State University}
  \country{USA}
}
\email{bgulmez@iastate.edu}

\begin{abstract}

Self-modifying code (SMC) allows programs to alter their own instructions, optimizing performance and functionality on x86 processors. Despite its benefits, SMC introduces unique microarchitectural behaviors that can be exploited for malicious purposes. In this paper, we explore the security implications of SMC by examining how specific x86 instructions affecting instruction cache lines lead to measurable timing discrepancies between cache hits and misses. These discrepancies facilitate refined cache attacks, making them less noisy and more effective. We introduce novel attack techniques that leverage these timing variations to enhance existing methods such as Prime+Probe and Flush+Reload. Our advanced techniques allow adversaries to more precisely attack cryptographic keys and create covert channels akin to Spectre across various x86 platforms. Finally, we propose a dynamic detection methodology utilizing hardware performance counters to mitigate these enhanced threats.

\end{abstract}
\maketitle

\section{Introduction}

Modern processors achieve enhanced performance through complex microarchitectures, incorporating features like sophisticated optimizations, more comprehensive pipelines, and advanced cache mechanisms. While these innovations boost efficiency, they also expose new vulnerabilities, evidenced by numerous attacks on shared microarchitectural components: cache~\cite{irazoqui2014wait,gruss2016flush+,zhang2012cross}, branch prediction unit~\cite{evtyushkin2018branchscope}, and translation look-aside buffer~\cite{gras2018translation}. These vulnerabilities have been exploited to bypass security measures and siphon sensitive data. Moreover, transient execution attacks leverage speculative and out-of-order execution to access transiently available data, exacerbating security risks~\cite{kocher2020spectre,lipp2020meltdown}. 

Several optimizations are present in the front-end of x86 processors, in which instructions are fetched and decoded into micro-operations ($\mu$ops) and transferred to the execution ports. Notably, the prefetching mechanism in the front-end unit fetches instructions even further in the pipeline speculatively to keep up with the speed of execution ports, assuming these instructions remain unaltered. However, if an application has write access to its instruction memory, any instruction can be overwritten dynamically during the code execution~\cite{ansel2011language}. This modification is detected by processors to maintain the correct flow of the instruction stream. x86 systems detect these changes through special hardware units and check mechanisms in the front-end, and all instructions after the modifying instructions are invalidated, causing a pipeline flush~\cite{kyker2003method,self_amd}, which is frequently referred to as self-modifying code (SMC) detection mechanism.

Resource sharing has been promoted on modern CPUs via optimizations such as Simultaneous Multithreading (SMT)~\cite{tullsen1995simultaneous}.
SMT technologies, i.e., Intel hyper-threading, enable multiple virtual processors that run disjoint tasks to share resources on the same core with up to 30\% performance gain with the same CPU die size~\cite{intelHTPerf}.
On the other hand, SMT introduces new attack vectors and security challenges.
Attacks such as PortSmash~\cite{aldaya2019port} are only feasible due to the real-time sharing of core-private resources, which otherwise were impractical. 
In May 2019, a class of Meltdown-style attacks collectively referred to as microarchitectural data sampling (MDS)~\cite{schwarz2019zombieload,van2019ridl}, 
also highlighted the negative impact of Intel hyperthreading on the security of modern CPUs.
However, the performance gained by SMT proved indispensable, outweighing any data leakage across logical processors.
These attacks can be even more applicable in Function-as-a-service (FaaS) cloud platforms with core multi-tenancy, such as Cloudflare Workers and Amazon Lambda instances.

When the processor encounters a self-modifying code snippet, the detection and correction mechanism slows down all threads running on the same physical core. Even though self-modifying code mechanism has been exploited to either degrade the performance of a processor through frequent pipeline 
flushes~\cite{aldaya2022hyperdegrade} or increase the speculative window size~\cite{ragab2021rage}, there is no extensive investigation on how self-modifying code as a standalone attack vector can directly leak private information such as cryptographic keys, unauthorized secrets in memory, and keystrokes while creating a high bandwidth covert communication. 
This paper delves into how attackers can exploit the SMC mechanism on x86 systems and leverage it to create high-resolution, less noisy microarchitecture side-channel attacks. 
Our study identifies a diverse set of x86 instructions, triggering the SMC-detection mechanism and extending the well-known Prime+Probe and Flush+Reload attacks to create covert channels on the L1 instruction (L1i) cache across SMT threads in the same physical core. SMaCk offers a more efficient cache attack tool for adversaries as cache hits and misses can be distinguished more robustly compared to traditional Prime+Probe attacks on the L1i cache~\cite{aciiccmez2007yet,aciiccmez2010new}. 

In summary, this paper:

\begin{itemize}
    \item explores which x86 instructions trigger the SMC detection mechanism on Intel and AMD processors. 
    \item systematically analyzes how the SMC detection mechanism affects front-end and back-end units in x86 processors by utilizing performance counters.    
    \item crafts new Prime+iProbe and Flush+iReload attacks on the L1 instruction cache to create covert channels between two SMT threads with a low error rate.
    \item shows that attackers can identify vulnerable cryptographic libraries by monitoring L1 instruction cache sets and leak a 2048-bit RSA key from a vulnerable cryptographic implementation.
    \item demonstrates an SMC-based single-trace side-channel attack on the Secure Remote Password (SRP) protocol in the most recent OpenSSL implementation by leveraging the L1 instruction cache, successfully extracting the server's private key exponent bits. 
    \item leverages the L1 instruction cache side-channel to demonstrate Spectre-type attacks, showing the inefficiency of shared data cache-based Spectre detection mechanisms.
    \item designs a performance counter-based defense technique to detect newly introduced SMC-based attacks on modern processors.
\end{itemize}

\section{Background}\label{sec:background}
Modern CPUs comprise multiple cores, a shared last-level cache (LLC), and a coherent memory subsystem. 
The core is designed based on a multi-stage pipeline in which operations are highly synchronized by the use of out-of-order and speculative execution techniques. We provide a brief background on different microarchitecture components.

\noindent\textbf{Front-end Unit} fetches program instructions from the instruction cache and places them into a queue. 
In a complex instruction set architecture, each instruction is first decoded into smaller micro-operations ($\mu$OPs). 
The decoded $\mu$OPs may be cached in the $\mu$OP cache to be reused. 
The core fetches a long sequence of instructions ahead of time, 
but it also fetches instructions that are not sequentially available in the program with the help of the branch predictor.
The branch predictor maintains the history of target branches and decisions to facilitate speculative fetching and execution of instructions from latent control flows.

\noindent\textbf{Back-end Unit} retrieves the $\mu$OPs from the allocation queue and then allocates resources while the instructions are waiting in the reorder buffer (ROB) to be completed.
In parallel, the scheduler sends the $\mu$OPs to various execution units depending on the availability of resources.
$\mu$OPs completed correctly will be retired by obeying the correct ordering of the program's instruction. 
If an error is detected in the ROB for an operation, the wrong outcome will be discarded, and corresponding $\mu$OPs will be rescheduled. 

\noindent\textbf{Memory Subsystem} consists of multiple levels of caches and buffers. The first level cache, L1 cache, is the fastest and smallest compared to other levels. 
It is separated into data (L1d) and instruction (L1i) caches where there is no interference between each other.
While the L1d cache keeps the data cache lines, the L1i cache stores the instruction cache lines. Intel processors have 64 sets for each L1 cache, and the number of ways is 8.
The second-level (L2) cache keeps both data and instruction cache lines. 
It is relatively slower than the L1 cache, but the L2 cache has a much higher volume. The last-level cache (LLC) is shared among all the physical cores, leading to a slower cache access time while having a larger capacity.

\noindent\textbf{Self-Modifying Code (SMC)} refers to programs that dynamically alter their instructions while they are executing~\cite{cai2007certified}. Traditionally, SMC has been used to optimize performance and manage resources efficiently, which requires high flexibility. SMC introduces complexity into the pipeline process because it changes the execution path. For instance, the fetch unit prefetches instructions stored in the L1i cache to the pipeline while enhancing CPU performance. However, when instructions modify themselves, this can lead to incorrect program behavior or system crashes. To handle this, processors must utilize various methods and apparatuses~\cite{zaidi2001system,murty2002apparatus,kyker2003method,botacin2020self} to detect SMC behaviors when instruction modifies the code. Once the processor detects SMC behavior, it flushes their entire pipeline, degrading the performance of all running tasks in the same core. Consequently, SMC conflicts are tied to the L1i cache because the SMC detection mechanism monitors the L1i cache region. The L1i cache is part of the front-end, whereas the L2 cache and LLC aren't in the front-end but in the back-end. Therefore, if SMC behavior modifies an instruction, the modification is only detected in the L1i cache.
\section{Threat Model}\label{sec:Threat Model}
In our threat model, we consider unprivileged attackers, targeting a victim running in the sibling thread of the same physical core through hyper-threading to leak sensitive information such as private keys, passwords, or secrets in the victim's memory. The processor pipeline is shared between two threads with microarchitectural units (L1i cache, TLB, execution ports) and the attacker can create SMC snippets with the available instructions in the targeted microarchitecture. We assume that the victim system has no exploitable vulnerabilities except the ones described hereafter.

\section{SMC ATTACK PRIMITIVES}\label{sec:SMCAttack}

Self-modifying code (SMC) has been used to slow down the execution of instructions to leak TLS-DH key exchanges~\cite{aldaya2022hyperdegrade} and secret characters through speculative execution~\cite{ragab2021rage}. 
While most of the experiments primarily focus on a thread's effect on its own execution, our approach enables one thread to affect its sibling thread in the same physical core.
The combination of SMC conflicts and a side-channel attack strengthens the signal-to-noise ratio by either increasing attack resolution or extending speculative window size. 

In our study, we propose to leverage SMC conflicts to create novel cache side-channel attacks within the instruction cache used in x86 systems. First, we show that an attacker can create an SMC conflict with different instructions rather than only using the \texttt{store}
instructions as exemplified in~\cite{ragab2021rage}. Next, we demonstrate that the time it takes to execute the SMC creating instruction provides enough granularity to distinguish L1i cache hit and miss, which can be used by an attacker to perform new variants of Prime+Probe and Flush+Reload attacks. Finally, we reverse engineer the behavior of different SMC-creating instructions with performance counters to analyze the root cause of slow-down in both the front-end and back-end of Intel and AMD processors.

\subsection{Analyzing SMC Conflict with Various x86 Instructions}
Modern x86 processors detect SMC occurrences during runtime with a sophisticated mechanism~\cite{kyker2003method,self_amd}. 
The main reason for the SMC conflict is the invalidation of cache lines already fetched speculatively due to the aggressive prefetching mechanism.
Hence, any instruction invalidating the fetched cache line could trigger the SMC mechanism. 
As a result, the execution pipeline, as well as the fetched instructions, are flushed to revert back to the immediate instruction after the conflict-creating instruction. 
In this section, we analyze different x86 instructions that could create SMC conflicts and measure the time it takes to execute those instructions. 

\noindent\textbf{Step 1: Preparing the L1i Cache.}
In our first step, we create a cache line (referred to as oracle) with read, write, and execute permissions. 
The cache line in this memory space is filled with 63 \textit{nop} instructions and 1 \textit{ret} instruction such that the instruction pointer returns to the caller function after the \textit{nop} instructions are executed as given in Listing~\ref{lst:l1i_prepare}. 
The TLB page for the oracle is created by storing a \textit{nop} instruction in the first byte to avoid any page walk that would affect the timing measurements. 
The same cache line is also flushed from the instruction cache. 
Next, the \textit{nop} instructions in the oracle are executed to validate the cache line and placed into the L1i cache. 
Finally, we place the \texttt{mfence} instruction to finish all the memory operations before the next instruction. 
These steps are required to load the cache line to the L1i cache, as this cache line will be modified in the next step to test whether different instructions create the SMC conflict.

\begin{lstlisting}[caption={L1 Instruction Cache State Preparation},language={[x86masm]Assembler}, label=lst:l1i_prepare, ,label=lst:l1i_prepare,basicstyle=\ttfamily\scriptsize, basewidth={.48em}, backgroundcolor=\color{white}, morekeywords={.data,.global,.quad,.space,.macro,.endm,mfence,rdtsc,shl,add,sub,pop,push,ret,lea,.text,.align,.rept,.endr}]
.align 0x1000
oracle_code_page:
  .rept 63
        nop
  .endr
  ret
movb $0x90, (uint64_t *)oracle_code_page 
clflush (uint64_t *)oracle_code_page
oracle_code_page()                       
mfence

\end{lstlisting}

\begin{figure}[t]
    \centering
    {
    
    \includegraphics[width=\linewidth]{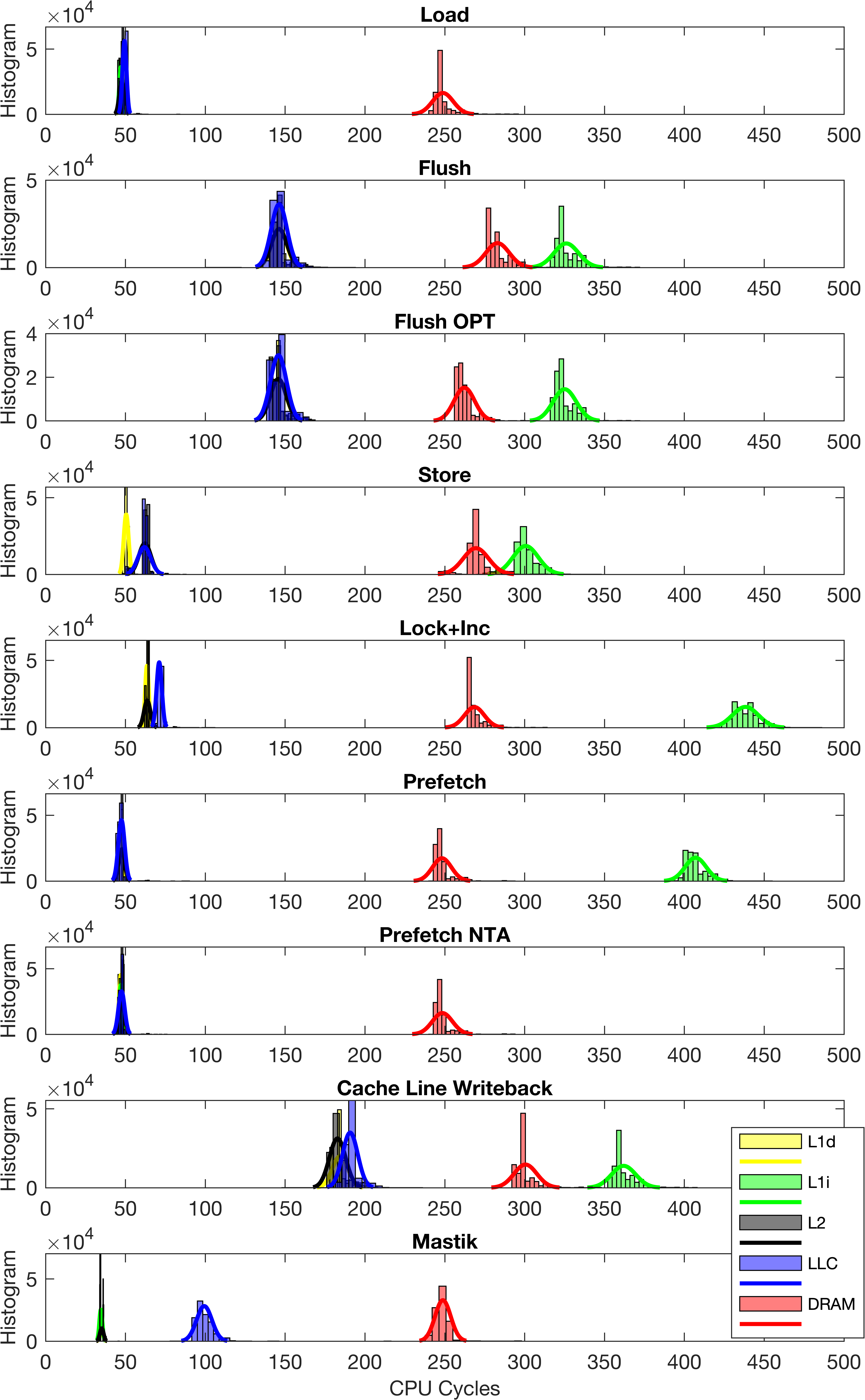}
    }
    \caption{CPU cycle time difference for various Probe strategies and Mastik~\cite{yarom2016mastik} method on Intel Cascade Lake microarchitecture.}
    \Description{CPU cycle time difference for various Probe strategies and Mastik~\cite{yarom2016mastik} method on Intel Cascade Lake microarchitecture.}
    \label{fig:CPUcycleTime_cascade}
\end{figure}

\noindent\textbf{Step 2: Measuring the Execution Time for SMC-creating Instructions.} In this step, our purpose is to determine whether an instruction that modifies the L1i cache line could cause an SMC conflict or not. 
Our hypothesis is that if an instruction creates an SMC conflict, the execution time for that instruction will increase compared to executing the same instruction on a memory address that is not in the L1i cache. 
We tested nine different instructions to create the SMC conflict and measured the execution time of each instruction as given in Listing~\ref{lst:SMC_inst}.
Each option from Line 3 to Line 12 is tested separately.
In Listing~\ref{lst:SMC_inst}, the \texttt{rdi} operand is the base address of the oracle. Each instruction is tested with five different microarchitecture states for 10,000 times: the oracle code page address is in the L1i cache, L1d cache, L2 cache, LLC, and DRAM.  
The execution times for each instruction in the Cascade Lake microarchitecture is given in Figure~\ref{fig:CPUcycleTime_cascade}.

\lstset{style=mystyle}
\begin{lstlisting}[caption={Measuring the execution time of nine different x86 instructions. These instructions represent load, flush, store, lock, prefetch, execute, and cache line write-back operations.},language=C,label=lst:SMC_inst,basicstyle=\ttfamily\scriptsize, basewidth={.48em}, backgroundcolor=\color{white},morekeywords={.data,.global,.quad,.space,.macro,.endm,mfence,rdtsc,shl,add,sub,pop,push,ret,lea,.text,.align,.rept,.endr}]
mfence
rdtsc
(1) mov (%rdi), %rax      // Load operation
(2) clflush (%rdi)        // Clflush operation
(3) clflushopt (%rdi)     // Clflushopt operation
(4) movb $0x90, (%rdi)    // Store operation
(5) lock                  // Lock+Inc operation  
    incb (%rdi)
(6) prefetch (%rdi)       // Prefetch operation
(7) prefetchnta (%rdi)    // Prefetchnta operation
(8) call oracle_code_page // Execute operation
(9) clwb (%rdi)           // Cache line write back operation
mfence
rdtsc
\end{lstlisting}

The timing values indicate that \texttt{clflush, clflushopt, store, lock, prefetch, clwb} instructions create a distinguishable time difference between L1i cache hit and miss events. 
These instructions are also evaluated with performance counters (MACHINE\_CLEARS:SMC) to verify that a diverse set of x86 processors triggers the machine clear event with the SMC conflict.

\noindent\textbf{Flush Instructions.} In the Cascade Lake microarchitecture, \texttt{clflush} and \texttt{clflushopt} instructions are executed in around 350 cycles if the memory operand is in the L1i cache. The timing value is considerably higher than other microarchitecture states such as LLC hit and DRAM access. Both flush instructions invalidate the cache line in the L1i cache, ensuring that future accesses to that memory address result in fresh data being loaded from the main memory. However, the instructions after the flush instruction may have already been fetched and executed in the back-end. Hence, both the front-end and back-end are flushed, leading to higher timing measurements. We observed that the time difference between the L1i cache hit and the LLC hit is more than 150 cycles. 
Similarly, flush instructions create 200 cycles difference between DRAM and L1i cache on AMD Ryzen 5 while the time difference reaches up to 300 cycles between L1i cache and LLC. The time difference between L1i cache, LLC, and DRAM states shows that both flush operations can be used to perform Prime+Probe and Flush+Reload attacks on AMD and Intel processors as described in Section~\ref{sec:casestudies}.

\noindent\textbf{Store Instruction.} The store instruction sets the dirty bit for the L1i cache line by updating its content, triggering the cache coherency mechanism to handle the inconsistency across the system's caches. We observed that even though the updated instruction is not executed, the SMC conflict is still created by a single update. The time it takes to execute the store operation is around 300 cycles, which is around 200 cycles more than the LLC hit. The time difference between DRAM and L1i cache is lower than other instructions, leading to 20 cycles difference. 
We also observed a significant time difference between the L1i cache hit and the LLC hit in the AMD Ryzen 5 processor with 150 cycles. While the time difference between L1i cache hit and DRAM access is close to each other, their execution time is still distinguishable with the \textit{rdtsc} instruction.

\noindent\textbf{Lock + Inc Instruction.} The \texttt{lock} instruction with the \texttt{incb} instruction modifies the \textit{nop} instruction stored in the oracle code page. Since this operation is executed atomically, the CPU detects this change on the executable instruction cache line as an SMC scenario. The execution time for the two instructions increases when the modified address is in the L1i cache, leading to 425 clock cycles on average. This pair of instructions has the highest execution time compared to other SMC-creating instructions, which is also consistent in other microarchitectures. While the time difference between the L1i cache and LLC hit is around 350 cycles, the time difference between the L1i cache and DRAM is around 150 cycles. 
In the AMD Ryzen 5 processor, all cache states are observable, and higher L1i cache hit cycle time shows that lock instruction creates an SMC conflict on AMD processors.

\noindent\textbf{Prefetch Instructions.} The \texttt{prefetch} instruction prefetches the content from the specified address and fetches it to the cache. 
If the memory address is already in the L1i cache and another thread modifies the L1i cache, it violates cache coherence mechanisms and causes core squash.
When a different instruction is fetched to that cache line from memory, the cache line invalidation triggers SMC conflicts.
This invalidation process triggers the SMC conflict. We observe that the time difference between L1i cache hit and LLC hit is around 350 cycles while the time difference between DRAM and L1i cache hit is 150 cycles. Interestingly, an SMC conflict is not triggered when the \texttt{prefetchnta} instruction is executed in the Cascade Lake microarchitecture. 
In AMD processors, we observed no time difference between the L1i cache hit and the LLC hit, concluding that prefetch instructions have no effect on creating SMC conflicts on these processors.

\noindent\textbf{Clwb Instruction.} The cache line write back (\texttt{clwb}) instruction writes back the content of the cache line to the memory that contains the virtual address specified with the memory operand from any level of the cache hierarchy in the cache coherence domain. After this instruction, the cache line usually stays in the cache to avoid a cache miss on a subsequent access. However, hardware may choose to invalidate the line from the cache hierarchy. Since the state of the cache line changes with the \texttt{clwb} instruction, the processor triggers the machine clear operation. We observed that the time difference between the L1i cache hit and the LLC hit is around 170 cycles. The time difference between DRAM access and L1i cache hit is close to 100 cycles. 
In contrast, the same instruction is not treated as an SMC conflict in the AMD Ryzen 5 processor.

\noindent\textbf{Comparison with Mastik~\cite{yarom2016mastik} L1i cache attack.} The Mastik toolkit~\cite{yarom2016mastik} provides an L1i cache attack by utilizing the Prime+Probe technique. We reproduced the access time histogram (last row in Figure~\ref{fig:CPUcycleTime_cascade}) for cache levels and memory with the Mastik tool. We observe that the L1i cache incurs an average of 34 cycles, and the L2 cache takes an average of 35 cycles with the Prime+Probe attack implemented in the Mastik tool. Since the time difference between L1i and L2 caches is only 1-2 cycles, it is challenging to distinguish cache evictions due to applications running in the neighbor virtual core. The low time difference in the Mastik tool leads to more noisy measurements, as discussed further in the single-trace side-channel attack in Section~\ref{subsec:case3}. Furthermore, the time difference between the L1i cache and L3 cache is around 60 cycles, while Lock+Inc SMC version reaches up to 350 cycles time difference for the same cache levels.

\begin{figure*}[ht]
    \centering
    {
    \includegraphics[width=\linewidth]{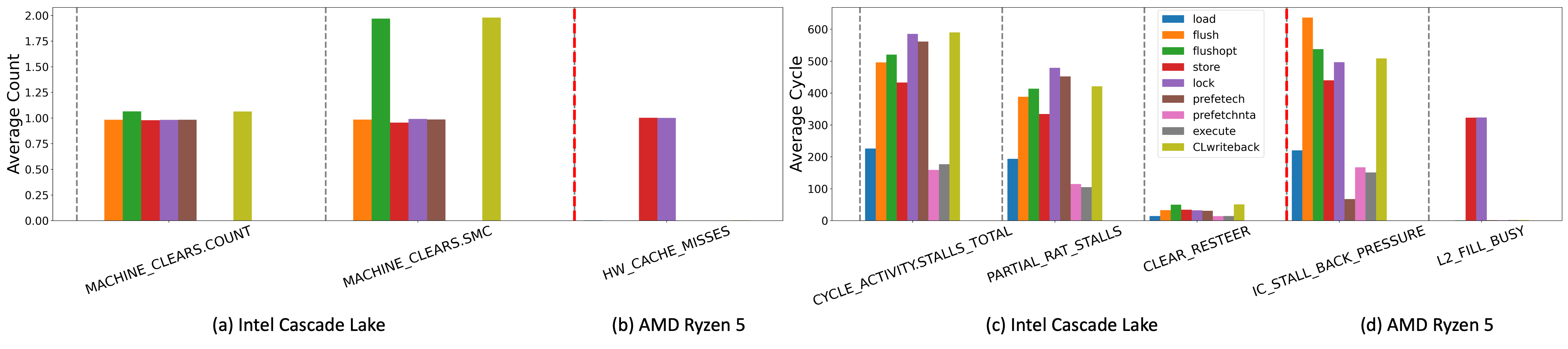}
    }
    
    \caption{Reverse Engineering SMC behavior on (a), (c) Intel Cascade Lake and (b), (d) AMD Ryzen 5 microarchitectures by using the PAPI~\cite{PAPI} tool. The counters belonging to Intel and AMD processors are separated by the red line.}
    \Description{Reverse Engineering SMC behavior on (a), (c) Intel Cascade Lake and (b), (d) AMD Ryzen 5 microarchitectures by using the PAPI~\cite{PAPI} tool. The counters belonging to Intel and AMD processors are separated by the red line.}
    \label{fig:counter_values}
\end{figure*}

\subsection{Reverse Engineering SMC Behavior on x86}\label{sec:reverse}

We use hardware performance counters to analyze how x86 processors behave when an SMC conflict occurs. We leverage the Performance Application Programming Interface (PAPI)~\cite{PAPI} to track hardware events at the microarchitecture level with minimal noise. We determined several hardware events on Intel and AMD processors, providing more detailed insight into the characteristics of SMC conflict resolution as given in Figure~\ref{fig:counter_values}. 
We start the counter before the SMC-creating instruction and stop the counter immediately after the instruction 10,000 times and take the average.
We separately perform the reverse engineering efforts on Intel and AMD devices as available performance counters are different in these processors.

\noindent \textbf{Reverse Engineering Intel x86.} We perform our experiments on the Intel Cascade Lake microarchitecture. The \texttt{MACHINE\_CLEARS.COUNT} event counts the number of any type of machine clears occurring during an execution. This counter confirms that six instructions trigger the machine clear mechanism while \texttt{MACHINE\_CLEARS.SMC} indicates that \texttt{clflushopt} and \texttt{clwb} instructions trigger the SMC counter twice. We believe that this counter is not working accurately for all the SMC cases. 

In the second part, we analyze the time spent on resolving the SMC conflict. First, we monitor the \texttt{CYCLE\_\allowbreak ACTIVITY.\allowbreak STALLS\_\allowbreak TOTAL} counter. The total stall cycle values are the highest for \texttt{lock} and \texttt{clwb} instructions (up tp 580 cycles), aligning with our \texttt{rdtsc} measurements. 
Next, we analyze the number of cycles spent at the front-end to flush the decoded stream buffer (DSB) and other instructions currently fetched/decoded. The \texttt{FRONTEND\_RETIRED:IDQ\_4\_BUBBLES}  event counts the cycles in which the front-end did not deliver any $\mu$ops (4 bubbles) for a period determined by the \textit{fe\_thres} modifier and which was not interrupted by a back-end stall. 
We gradually increased the threshold and noticed that all instructions created 30 stall cycles on the front-end. 
When we check the \texttt{INT\_MISC.CLEAR\_RESTEER\_CYCLES} counter, we observed that the processor waits for around 35-40 cycles to issue the new instructions at the back-end after the machine clears. We expect that new instructions take the legacy path and go through the fetch and decode stage, leading to additional cycles to reach the back-end again. 

The SMC conflict behaves like a fence instruction by serializing the instructions after the conflict is detected. For the back-end stall, we leverage the \texttt{PARTIAL\_RAT\_\allowbreak STALLS.\allowbreak SCOREBOARD} counter, reporting the number of cycles the CPU issue-pipeline was stalled due to serializing operations. The counter value is monitored with and without the attack instruction, and we observed that the processor spends around 200 cycles to complete the serialization in the back-end for the \texttt{store} instruction after the counter overhead is removed (around 100 cycles) in Figure~\ref{fig:counter_values}.
All in all, Intel processors spend more cycles in the back-end to revert back to the correct order of instructions.

\noindent \textbf{Reverse Engineering AMD x86.} AMD processors provide a different set of counters compared to Intel processors. For example, AMD architectures have no option to observe machine clears. Instead, we focus on the instruction cache and pipeline stall-related counters to observe the effects of our SMC code snippets. We observed that \texttt{clflush}, \texttt{clflushopt}, \texttt{store}, and \texttt{lock} instructions create distinguishable time differences between L1i cache and LLC accesses. The time difference mainly comes from the back-end unit as the \texttt{INSTRUCTION\_PIPE\_STALL:BACK\_PRESSURE} counter increases significantly. We observed that the counter reports around 500 cycles of stall occurring in the back-end unit for \texttt{clflush} operations, which matches with the timing values. The effect of SMC conflict on the instruction cache is also verified with the \texttt{INSTRUCTION\_\allowbreak CACHE\_\allowbreak LINES\_\allowbreak INVALIDATED:FILL\_\allowbreak INVALIDATED} counter. As expected, the counter value increases by one for the SMC-creating instruction, stating that the cache line is invalidated with the \texttt{store} and \texttt{lock} instructions. Then, the invalidated cache line is requested from the L2 cache as the \texttt{CYCLES\_\allowbreak WITH\_\allowbreak FILL\_\allowbreak PENDING\_\allowbreak FROM\_\allowbreak L2:L2\_\allowbreak FILL\_\allowbreak BUSY} counter increments with the same amount we observed from the back-end counter, which shows that requesting cache lines from L2 cache is considered as back-end stall in AMD counters. For \texttt{clflush} and \texttt{clflushopt} operations, the counter has no increase because these operations do not bring the cache line to the cache. Due to the lack of precise counters on AMD processors, we identified a few counters to analyze how AMD processors handle SMC conflicts.

\noindent \textbf{SMC Root Cause Analysis.} 
%
Intel patent on the SMC detection mechanism describes that the instruction before the SMC-creating instruction is the last committed one to the register file, while all the prefetched instructions after the SMC-creating instruction are flushed from the pipeline~\cite{intel_smc}. Then, the modified instruction cache line is fetched from memory to fix the memory inconsistency. Hence, the instructions after the SMC-creating instruction are still executed in an out-of-order way because the new instruction is not determined yet due to the high access time from the main memory. This hypothesis is also supported by Ragab et al.~\cite{ragab2021rage} as they show that speculative loads are still executed in their test case (Pg. 6 Listing 1) with the increasing transient window size.
We also observed the same high timings for SMC-creating instructions when \textit{mfence} instructions are not placed in the code snippet (Listing~\ref{lst:SMC_inst}), indicating that timing measurements do not rely on the delayed memory operations. Since the fetch, decode, and execute pipelines are shared between two sibling threads in the same physical core, and these pipelines are flushed when the SMC conflict is detected, the sibling thread is stalled for each SMC conflict detection. As a result, the observed high time differences for each SMC-creating instruction are due to the time spent on the executed instructions, and the main reason is the high instruction transfer time from memory to the L1i cache.
However, it is important to acknowledge that the SMC squash effect might also influence the measured delay. Specifically, if the SMC-creating instruction commits quickly before the pipeline is squashed, the timing utilizing the second \textit{rdtsc} timer (line 14 in Listing~\ref{lst:SMC_inst}) could reflect the delay and induce more timing discrepancy as \texttt{Store} and \texttt{Lock} instruction on Intel CPUs creates more delays as depicted in Figure~\ref{fig:CPUcycleTime_cascade}.

Our findings in Figure~\ref{fig:CPUcycleTime_cascade} also support this claim as we observe that transferring instructions from memory to the L1i cache takes 250 cycles without SMC conflict (a load operation implemented in the Mastik Tool). However, additional SMC detection mechanisms and pipeline flush processes introduce delays in L1i cache operations, resulting in more than 250 cycles for each instruction creating the SMC conflict (green bars in Figure~\ref{fig:CPUcycleTime_cascade}).

\noindent\textbf{Outcome.} In summary, invalidating a cache line in the L1i cache creates inconsistency issues between the memory and the L1i cache, even though the content of the instruction cache line may not be modified. The inconsistency is resolved by stopping instruction fetching and clearing the instructions already transferred to the execution ports and ROB. Both Intel and AMD processors spend more time flushing the back-end unit compared to the front-end while resolving the SMC conflict. Each processor demonstrates a different timing behavior for each instruction, leading to a unique behavior. We also observed that a single SMC conflict leads to a 235-cycle slowdown in the sibling thread.
\section{Case Studies }\label{sec:casestudies}
\subsection{Case Study I: Prime+iProbe and Flush+iReload Covert Channels with SMC}\label{subsec:case1}

In this case study, we exploit the SMC behavior on x86 processors to create a high bandwidth and low error rate covert channel utilizing Prime+Probe and Flush+Reload techniques. 

\noindent\textbf{Prime+iProbe.} We adopt the Prime+Probe approach to create our Prime+iProbe covert channel. First, the sender and the receiver agree on an L1i cache set to transmit the bits. Next, the sender fetches an instruction cache line from either the lower level cache (L2 or L3) or memory into the L1i cache set by executing an instruction if the transmitted bit is '1'. Otherwise, the sender performs a dummy function for a certain amount of time to send '0'. The receiver allocates a contiguous memory region and creates an eviction set with eight cache lines for the pre-determined set number by filling them with \textit{nop} instructions. During the prime phase, one \textit{nop} instruction is executed to place the cache lines into the L1i cache. In the probe phase, the receiver leverages one of the SMC primitives described in Section~\ref{sec:SMCAttack} to create the SMC conflict and measure the time for each cache line. If at least one of the cache lines produces a low execution time value, the sender transmits a '1' because the sender evicted one cache line belonging to the receiver, and no SMC conflict occurred for the evicted cache line. Otherwise, all cache lines will create SMC conflicts, leading to high \textit{rdtsc} timings. 

There are three parameters affecting the error rate and bandwidth: 1) \textit{The duration of dummy operations ($\tau_{d}$)} determines the length of the '0' bit. This duration should be long enough to ensure the receiver can observe the no-activity region in the cache set. 2) \textit{The number of loads ($N_l$)} for a single '1' bit transmission is crucial as the receiver may not be able to detect consecutive '1's. We prefer creating more than one load for a single bit to increase the chance of detecting consecutive '1's. 3) \textit{the waiting time between prime and probe phases ($\tau_{w}$)} is crucial to observe the activities in the L1i cache. If the receiver has a low waiting time in between, the probability of missing the L1i cache eviction caused by the sender increases gradually. In the meantime, more frequent pipeline flushes lead to slower sender time, eventually hurting the bandwidth. 

\begin{figure}[t!]
    \centering
    \includegraphics[width=\linewidth]{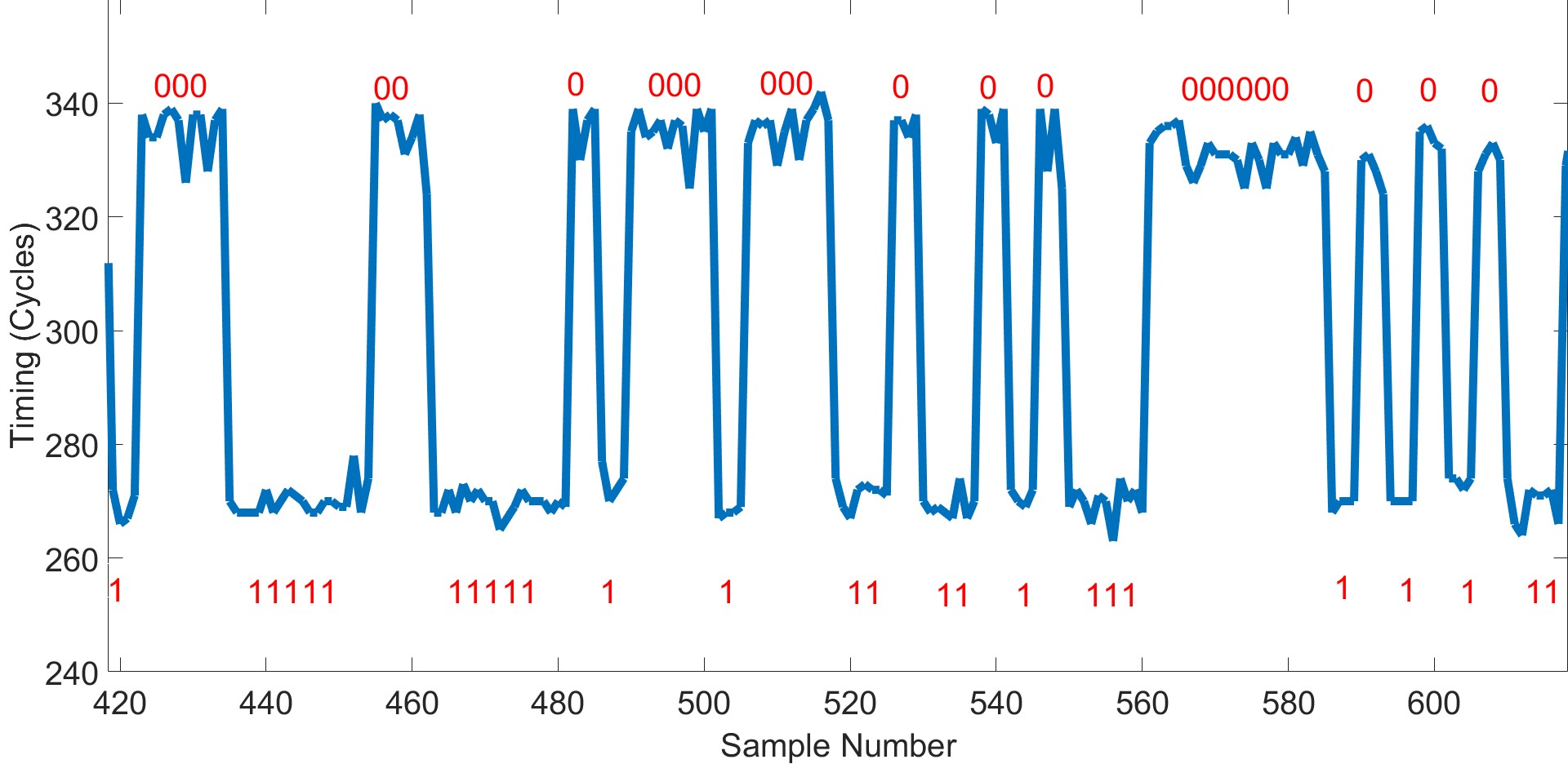}
    \Description{SMC timings measured by the receiver using Prime and LMS attack (blue line) and the automatically assigned bit values (red digits). The peaks for the 0s and 1s can be distinguished. This experiment is conducted in the Tiger Lake microarchitecture.}
    \caption{SMC timings measured by the receiver using Prime and LMS attack (blue line) and the automatically assigned bit values (red digits). The peaks for the 0s and 1s can be distinguished. This experiment is conducted in the Tiger Lake microarchitecture.}
    \label{fig:covert}
    
\end{figure}

\noindent\textbf{Evaluation.} Our test setup is based on the Intel Cascade Lake microarchitecture with a 32 KB instruction cache, 64 cache sets, and 8-way associativity. 
The duration of dummy operations for a single 0 bit is chosen as an empty for loop with $\tau_{d} = 10,000$ iterations. This duration leads to four samples per one '0', as shown in Figure~\ref{fig:covert}. 
Next, we determined that $N_l = 10$ loads create four consecutive low timings for the receiver. 
Finally, $\tau_{w} = 1,000$ iterations of an empty for loop is decided as the final value for the waiting time. This amount of waiting time is sufficient to detect the majority of load operations. As shown in Figure~\ref{fig:covert}, 0s and 1s can easily be transmitted between sender and receiver. The high time difference between low and high peaks also leads to a lower error rate in our covert channel, as shown in Table~\ref{tab:covert}. We achieve up to 189 Kbit/s with the Prime+iFlush covert channel, having only a 0.3\% error rate.

\begin{table}[h]
\small
\centering
\caption{SMC-based covert channels on instruction cache based on Flush+Reload and Prime+Probe attacks. The covert channels are created on the Intel Cascade Lake architecture. App. means applicability of the attack.}
\setlength{\tabcolsep}{5pt}
\scalebox{0.95}{
\begin{tabular}{l|c|c|c}
\textbf{Covert Channel} & \textbf{App.} & \textbf{Bit Rate (Kbit/s)} & \textbf{Error Rate (\%)} \\\hline
Prime+iFlush & $\checkmark$ & 188.9 & 0.3 \\\hline
Prime+iFlushopt & $\checkmark$ & 183.2 & 0.6 \\\hline
Prime+iLock & $\checkmark$ & 134.2 & 1.4 \\\hline
Prime+iPrefetch & $\checkmark$ & 120.1 & 1.2 \\\hline
Prime+iStore & $\checkmark$ & 123.2 & 0.6 \\\hline
Prime+iClwb & $\checkmark$ & 133.5 & 0.4 \\\hline
Flush+iFlush & $\checkmark$ & 660.2 & 0.9 \\\hline
Flush+iFlushopt & $\checkmark$ & 670.3 & 0.8\\\hline
Flush+iLock & $\times$ & N/A & N/A \\\hline
Flush+iPrefetch & $\checkmark$ & 452.7 & 0.4\\\hline
Prime+iStore & $\times$ & N/A & N/A \\\hline
FLush+iClwb & $\checkmark$ & 203.6 & 1.8\\\hline

\end{tabular}
}
\label{tab:covert}
\end{table}

\noindent\textbf{Flush+iReload.} We implement Flush+iReload covert channels based on the Flush+Reload technique. We create a shared code page filled with \textit{nop} instructions between two processes with read and execute permissions similar to a shared library structure in OS. This setup is similar to Flush+Reload attack scenarios in which the deduplication mechanism merges the same read-only code pages in the memory~\cite{irazoqui2014wait}. The sender executes an instruction in the shared code page to $N_l = 10$ times to send a single '1' bit. For the '0' bit, the sender executes an empty for loop with $\tau_d = 10,000$. The receiver executes SMC-creating instructions except \texttt{store} and \texttt{lock} as the code page has no write permission. 
As there is only one cache line used to transmit the bits, we can achieve a faster transmission rate with Flush+iReload channels.

\noindent\textbf{Evaluation.} We compare the covert channel bandwidth with various SMC-based covert channels as given in Table~\ref{tab:covert}. The Prime+iFlush side-channel achieves the highest bandwidth with 188.9 Kbit/s while creating a low error rate among Prime+iProbe techniques. Similarly, Prime+iFlushopt achieves a high bandwidth and low error rate. Prime+iLock leads to a higher error rate and lower bandwidth compared to flush-based SMC since the lock instruction creates a higher time difference, as given in Figure~\ref{fig:CPUcycleTime_cascade}, which results in spending more time probing the cache lines. 

Both Flush+iFlush and Flush+iFlushopt channels achieve around 660 Kbit/s transmission rate with less than 1\% error rate. The Flush+iPrefetch technique is comparably slower than flush-based covert channels since the \texttt{prefetch} instruction requires the \texttt{clflush} instruction to be executed before the cache line is brought into the L1i cache in order to create an SMC conflict. 
Finally, the \texttt{clwb} instruction requires more repetitions from the sender process to create distinguishable 0s and 1s. We tripled the number of executions and empty for loop size in the sender process to increase the signal-to-noise ratio, resulting in around 200 Kbit/s transmission rate. We also evaluated the Prime+iLock covert channel on the AMD Ryzen 5 microarchitecture. We observed that the covert channel is more noisy than on Intel Cascade Lake. One of the reasons is that we utilized the \textit{rdtsc} instruction to measure the time of executing lock and store instructions, which has 21 cycles resolution in our setup. Hence, the time difference between SMC and non-SMC behavior is less distinctive, leading to a bandwidth of 98.2 Kbit/s with a 2.5\% error rate.

\subsection{Case Study II: RSA Key Recovery}\label{subsec:case2}

In this case study, we describe how Prime+iProbe attacks can be utilized to perform an end-to-end side-channel attack. The experiments were performed on the Intel Tiger Lake microarchitecture to show that our attack is feasible in different microarchitectures.

\noindent\textbf{Step 1: Vulnerable Library Detection} In this part, we explore how the Prime+iStore attack can detect the targeted cryptographic library as well as its version. 
We assume that the victim uses one of the most popular libraries (OpenSSL or Libgcrypt) for the RSA decryption task while the attacker aims to identify the correct library and version. 

In the offline phase, we randomly selected 14 library versions from Libgcrypt and 20 versions from OpenSSL. 
We monitor all L1i cache sets in order, starting from set 0. We collect 100 Prime+iStore timings from each set, resulting in 6400 samples from 64 sets.
Next, we convert the timings to hit and miss based on the threshold (100 cycles for Tiger Lake). 
The number of activities in each set is summed up, and 64 activity values are created for each measurement. For each library version, we collected 8 measurements with varying decryption keys, resulting in 272 measurements in total. This dataset is used to train a k-th nearest neighbor (kNN) machine learning model. The model is trained with cross-validation, three closest neighbors, and Euclidean distance parameters, which achieved 100\% accuracy during the cross-validation process.

In the online phase, the attacker aims to detect the correct library version with only one measurement. We collected two measurements for each library and converted them to activity traces. The resulting feature vectors are then classified with the pre-trained model. We achieved 97\% accuracy in detecting the correct library. Note that there are slight differences between close library versions, making it more challenging to detect the correct library. However, the memory offsets for different functions used by each library differ, which creates a distinct fingerprint for each library. 

\begin{figure}[t]
    \centering
    {
    \includegraphics[width=\linewidth]{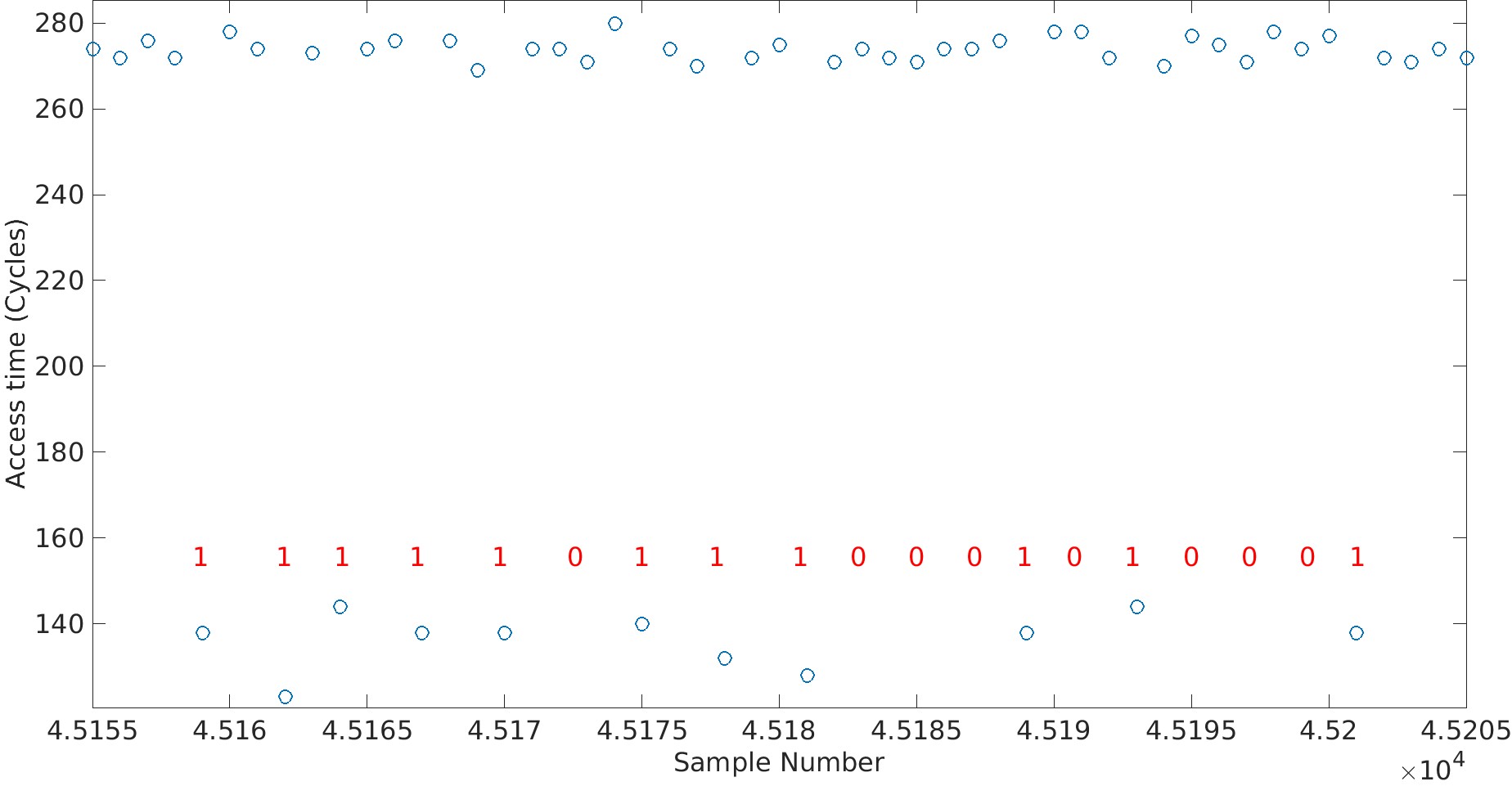}
    }
    \caption{Low timing values indicate that there is a multiplication activity as one of the cache lines is evicted from the L1i cache. The experiment is performed on Intel Tiger Lake with the Prime+iStore attack.}
    \Description{Low timing values indicate that there is a multiplication activity as one of the cache lines is evicted from the L1i cache. The experiment is performed on Intel Tiger Lake with the Prime+iStore attack.}
    \label{fig:RSA_bits}
\end{figure}

\noindent\textbf{Step 2: Multiplication Set Detection.}  It is crucial for an adversary to detect the L1i cache set used for multiplications. For this purpose, we profiled the Libgcrypt 1.5.1 RSA implementation. In the offline phase, we collect measurements consisting of 20,000 samples from each set with the Prime+IStore technique. 
The number of activities (cache misses) in each cache set is recorded to create the feature vector as depicted in Figure~\ref{fig:RSA_bits}. We collected 500 measurements from cache sets used for the multiplication operation and 500 more measurements from other cache sets. The dataset is divided into training (80\%) and test (20\%) datasets to train a binary classification kNN model. 
In the online phase, the attacker collects measurements from different sets and aims to detect the cache set used for multiplication. The kNN model can detect the correct set with a 96\% success rate. We further collected measurements from older versions of Libgcrypt, and the success rate remains the same. 

\begin{figure}[t]
    \centering
    {
    \includegraphics[width=\linewidth]{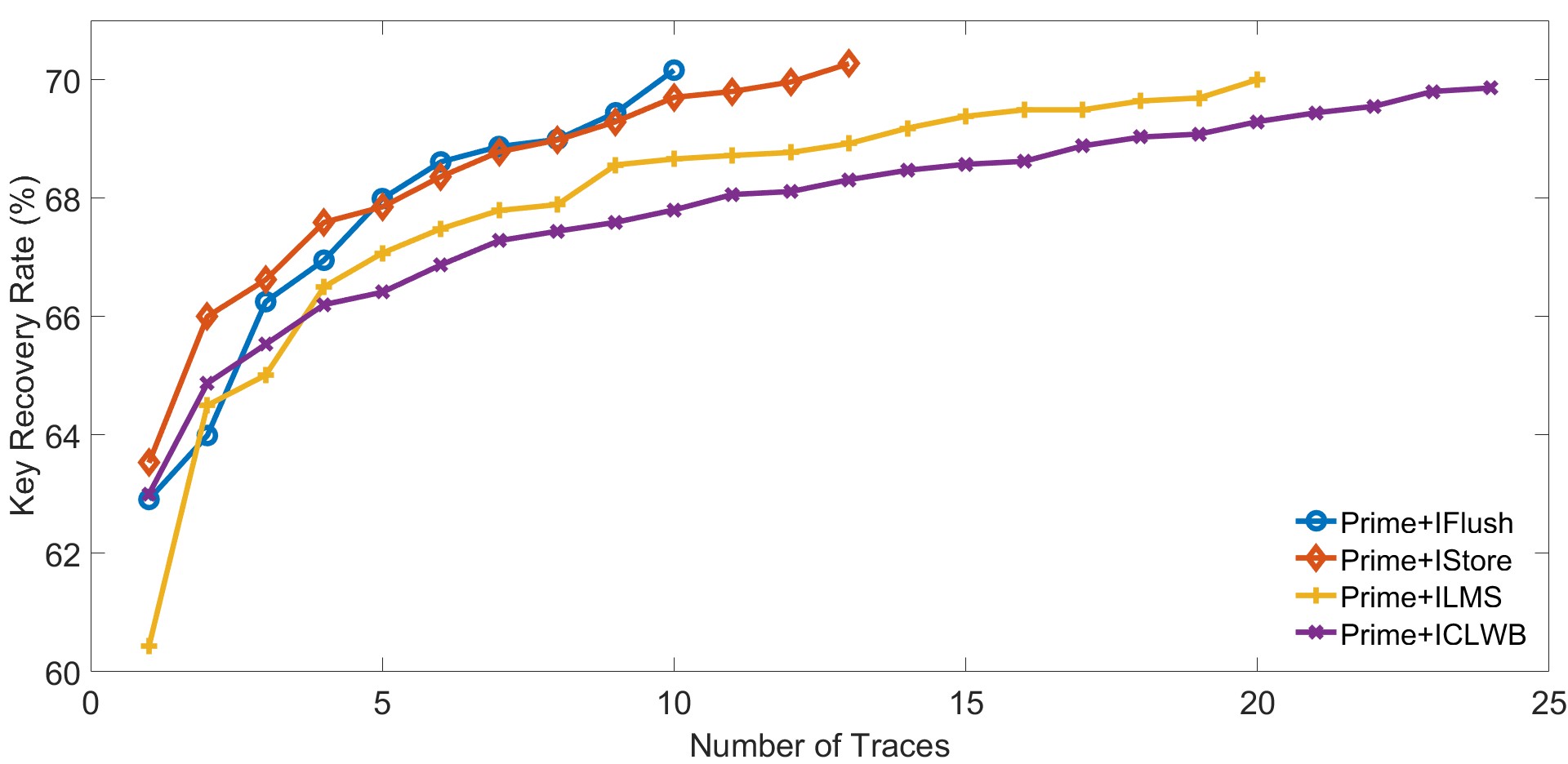}
    }
    \caption{The number of traces required to achieve 70\% key recovery rate for Flush, Store, Lock and Clwb-based SMC attacks on Intel Tiger Lake.}
    \Description{A graph showing the number of traces required to achieve a 70\% key recovery rate for Flush, Store, Lock, and Clwb-based SMC attacks on Intel Tiger Lake.}
    \label{fig:RSA_success_rate}
\end{figure}

\noindent\textbf{Step 3: RSA Key Recovery.} This step is implemented with several Prime+iProbe attacks to demonstrate the 2048-bit RSA decryption key recovery from the Libgcrypt 1.5.1 library by profiling multiplication operations. 

Our key recovery attack consists of several steps. In the first step, we decide the amount of time needed to be placed between the Prime and Probe steps. We identify both the slow-down amount on the RSA decryption process and the number of multiplication operations detected by the attack with different wait times. When the wait time is high, the RSA decryption process runs faster, limiting the attack's resolution and missing several multiplication function calls. If the wait time is very low, the RSA decryption runs very slowly, and multiplication function calls are less likely to be detected, leading to a smaller number of activities. We determined that an empty \textit{for} loop with 700 iterations between the Prime and Probe phases fulfilled our requirements. The time difference between multiplication operations increased by five times while one Prime+iProbe sample collection takes around 7000 cycles. 
We expect to have three Prime+iProbe samples without any activity if there are two multiplications consecutively ('11'). The number of no-activity samples is increased by 2 for each 0 bit between two consecutive multiplications. Hence, our attack can be performed on the RSA decryption process with a sufficient resolution.

In the second step, we collect Prime+iProbe samples while running the decryption process. When the multiplication function (\textit{mul\_n}) is called, at least one attacker-controlled cache line is evicted from the profiled cache set. When this cache line is invalidated with one of the SMC techniques, it takes less time to execute the specific operation since SMC conflict does not occur on that line. We collect multiple traces to recover a large portion of the 2048-bit RSA decryption key. As shown in Figure~\ref{fig:RSA_success_rate}, a single trace is successful in leaking 63\% of the RSA key when the Prime+iFlush attack is performed. As stated in previous works~\cite{yarom2014flush+,inci2015seriously}, recovering at least 70\% of the decryption bits is sufficient for attackers to leak the entire RSA decryption key. With the increasing number of traces, an attacker can reach up to 70\% secret key recovery while the number of required traces changes depending on the attack type. We observed that the Prime+iFlush attack only needs 10 traces to recover 70\% of the bits. Similarly, the Prime+iStore attack needs around 13 traces to recover 70\% of the bits correctly. On the other hand, Prime+iLock requires 20 traces to leak 70\% of the bits. We noticed that the relatively poor performance of the Prime+iLock attack is due to the high number of missed multiplication activities compared to other attacks. 
We also noticed that the \texttt{clwb} instruction takes a longer time, leading to more noisy samples as the time between prime and probe stages increases in parallel. 
Note that our attack does not assume that there is a shared memory page between the attacker and victim, which leads to a higher noise amount compared to Flush+Reload attacks. More interestingly, our Prime+Probe attack does not increase the last-level cache miss counters significantly, making it difficult for the current performance counter-based defense mechanisms to detect our attack as discussed in Section~\ref{sec:dynamic_detection}.

\subsection{Case Study III: OpenSSL SRP Single Trace Attack}\label{subsec:case3}
In this section, we leverage the Prime+iProbe attack to leak a secret key with only a single trace collected from the Secure Remote Password (SRP) protocol in OpenSSL. OpenSSL SRP protocol is used for secure password-based authentication between a client and server without transmitting the actual password, enhancing security against eavesdropping and replay attacks~\cite{wu1998secure,taylor2007using}. 

Both client and server compute a shared key to authenticate each other without sharing a password. In the registration phase, first, a publicly known group $G$ of prime order $p$ and a generator $g$ are created. Then, a hash function is used to compute a group related parameter $k=H(g\,||\,p)$. A verifier is calculated by the server v = $g^x\,mod\,p$ where $x = H(salt\,||\, H(client\_id\,||\,pwd))$. $pwd$ belongs to the client and the server does not have access to the $pwd$. Hence, the server keeps the password in the form $(client\_id,\,v,\,salt)$. In the login phase, the client generates a random number $a$ uniformly chosen from $(1,p-1)$, computes $A=g^a \,mod\,p$ and sends it to the server. Then, the server generates a secret random number $b$ chosen from $(1, p-1)$, computes $B=kv + g^b\,mod \,p$ and sends both the salt and $B$ to the client. Finally, the client recovers the verifier $v = g^x\,mod\,p$. If the secret exponent $b$ can be recovered by an attacker, a malicious server can be generated to authenticate the clients and access confidential information from the clients. Since $b$ is secret and randomly chosen for each authentication, an attacker needs to leak $b$ with a single trace, which is the target in our attack.
\begin{lstlisting}[caption={Implementation of the \texttt{SRP\_Calc\_server\_key} function in OpenSSL v1.1.1w SRP library.}, label=lst:SRP_library, language=Python, basicstyle=\ttfamily\scriptsize, basewidth={.48em}, backgroundcolor=\color{white}]
def SRP_Calc_server_key(A, v, u, b, N):
    BIGNUM *tmp = NULL, *S = NULL;
    BN_CTX *bn_ctx;
    
    # S = (Av^u)^b mod N
    BN_mod_exp(tmp, v, u, N, bn_ctx)
    BN_mod_mul(tmp, A, tmp, N, bn_ctx)
    
    S = BN_new();
    BN_mod_exp(S, tmp, b, N, bn_ctx) 
    
    return S; 
\end{lstlisting}

Even though the client implementation was exploited to leak passwords~\cite{de2021parasite} and then patched by setting the constant time flag within the client key generation function \texttt{SRP\_Calc\_client\_key}, the server-side key generation \texttt{SRP\_Calc\_server\_key} still remains unprotected against timing attacks as given in Listing~\ref{lst:SRP_library}. The omission of \texttt{BN\_\allowbreak FLG\_\allowbreak CONSTTIME} flag allows attackers to leak the server's private key exponent bits (\textit{b} in the Listing~\ref{lst:SRP_library}).

In the \texttt{SRP\_Calc\_server\_key} function, modular exponentiation is executed using the \texttt{BN\_mod\_exp\_mont} function to compute the server's shared secret key, $S=(A*v^u)^b\,mod\, N$. The algorithm processes the bits of the secret exponent \textit{p} (the binary representation of \textit{b}) from the most significant bit (MSB) to the least significant bit (LSB), using sliding window exponentiation. The square operation (\texttt{bn\_mul\_mont\_fixed\_top (r, r, r, mont, ctx)}) is executed if the \texttt{BN\_is\_bit\_set} function returns 0 (Line 3 and Line 12 in Listing~\ref{lst:BN_mod_exp_mont}). Conversely, if the bit is '1', the algorithm accumulates a window of bits with a maximum size of 6, \texttt{wvalue}, to reduce the number of multiplications while executing consecutive square operations for the number of bits in the window before the multiplication. Hence, the number of bits in each window leads to a variable timing from Line 9 to Line 20 in Listing~\ref{lst:BN_mod_exp_mont}. Our attack relies on the elapsed time between each execution of the jump instruction associated with the infinite loop in Line 2 since the operations (squares and multiplications) based on the secret (\textit{p}) bits lead to varying time differences. 
\begin{lstlisting}[caption={The implementation of the \texttt{BN\_mod\_exp\_mont} function in \texttt{/crypto/bn/bn\_exp.c}} OpenSSL-1.1.1w. The code has been modified as a Python-like syntax to save space., label=lst:BN_mod_exp_mont, language=Python, basicstyle=\ttfamily\scriptsize, basewidth={.48em}, backgroundcolor=\color{white}]
def BN_mod_exp_mont(rr, a, p, m, ctx, in_mont):
    while True:
        if not BN_is_bit_set(p, wstart):
        # Square
            if not bn_mul_mont_fixed_top(r, r, r, mont, ctx): 
            wstart -= 1
            continue
        # Window 
        for i in range(1, window):
            if wstart - i < 0:
                break
            if BN_is_bit_set(p, wstart - i):
                wvalue <<= (i - wend)
                wvalue |= 1
                wend = i
        # Square
        for _ in range(wend + 1):
            if not bn_mul_mont_fixed_top(r, r, r, mont, ctx): 
        # Multiply
        if not bn_mul_mont_fixed_top(r, r, val[wvalue >> 1], mont, ctx):
\end{lstlisting}

\noindent\textbf{Experiment Setup.} Our experiments were performed on an Intel Tiger Lake microarchitecture running Ubuntu 22.04 with 16GB RAM. We utilized OpenSSL 1.1.1w (up-to-date) to ensure our findings are relevant to current implementations. 

\noindent\textbf{Results.} Our attack relies on the Prime+iStore attack. Since the server calculates the shared key only once for each client, our purpose is to leak as many bits as we can in one trace and then recover the remaining bits as explained in~\cite{de2021parasite,yarom2014flush+,inci2015seriously}. The address of the "jump" instruction associated with the infinite for loop is determined through the disassembled version of the \texttt{libcrypto.so} binary to monitor the specific instruction cache set. Next, an authentication request to the server is sent to calculate the shared key. We assume that the server process and the attacker share the same physical core as described in Section~\ref{sec:Threat Model}. In the meantime, the attacker starts executing the Prime+IStore attack. 

The server's secret key ($b$) size is determined by the group size. While a larger group size provides better security, the incurred performance overhead increases in parallel. We tested our attack with four different group sizes: 1024, 2048, 4096, and 6144.
We manually set static values for the client's public ephemeral value (\textit{A}), username, password, and salt while the server's private keys (\textit{b}) are generated randomly.

The time difference between each iteration is 500-600 cycles for back-to-back square operations when two consecutive bits are "00" with a group size of 1024. When the Prime+IStore attack runs in parallel, the time difference increases to 2000 cycles due to frequent core resource flushes. This performance degradation enables the attacker to collect at least one sample between each square. When a sliding window operation is executed, the multiplication takes 6000 cycles and each additional bit in the window takes 2000 cycles. Hence, our attack has sufficient resolution to detect activities in each iteration even with the smallest group size. However, it is not possible to know which array entry is accessed for the window multiplication, leading to unknown bits in each window. If the window size is 6, the middle four bits "1XXXX1" are unknown. Based on our experiments, the randomly generated keys have around 45\% unknown bits that cannot be leaked through the instruction cache channel. Still, an attacker can perform dictionary search and mathematical algorithms to leak the entire key~\cite{inci2015seriously,de2021parasite}. 

\begin{figure}[t]
    \centering
    \includegraphics[width=\linewidth]{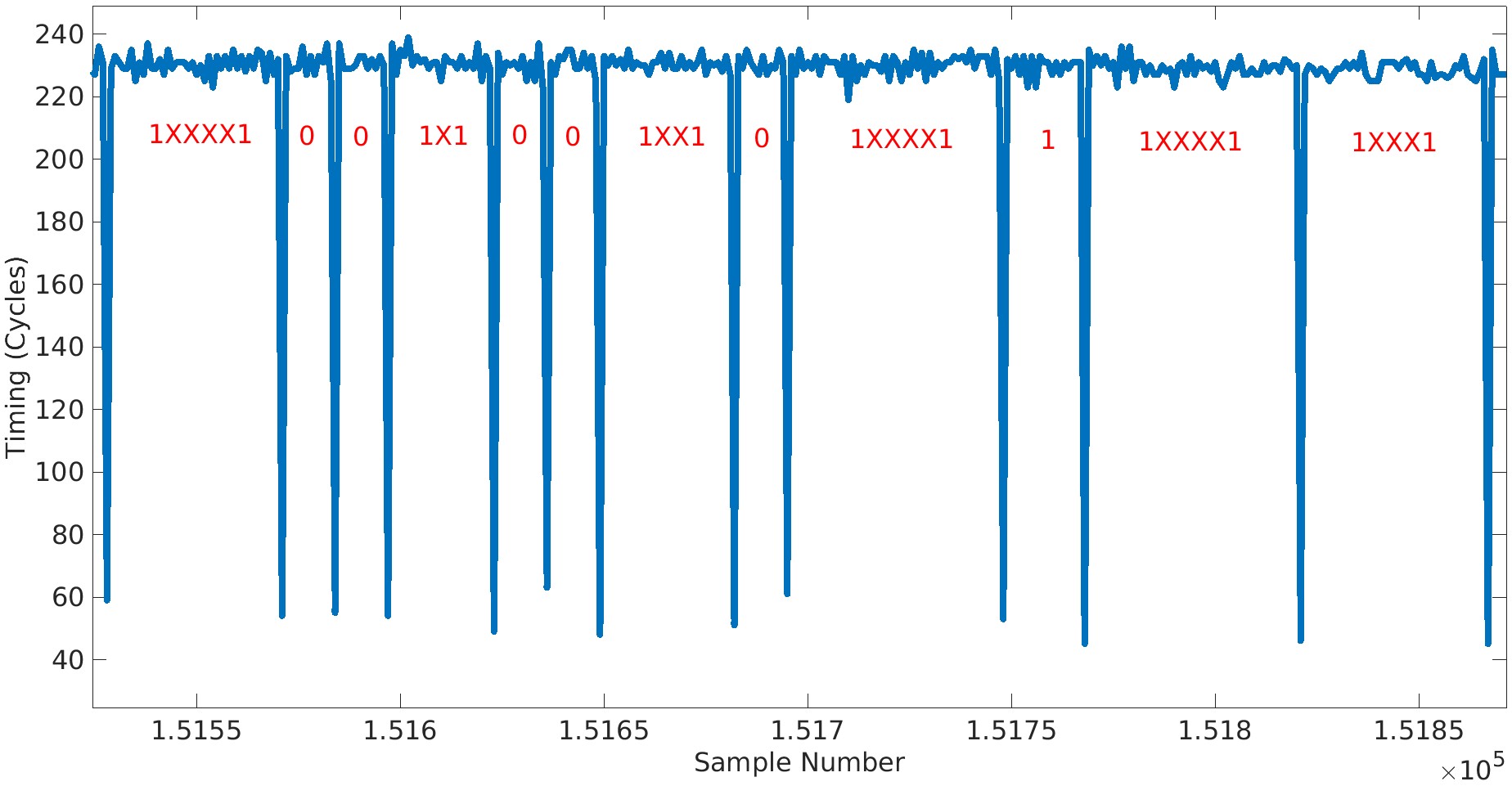}
    
    \caption{The cache activity monitored by the Prime+IStore attack during the secret exponent computation with the 6144 group size. "X" bits represent the unknown bits in the window multiplicand.}
    \Description{The cache activity monitored by the Prime+IStore attack during the secret exponent computation with the 6144 group size. "X" bits represent the unknown bits in the window multiplicand.}
    \label{fig:SRP}
    
\end{figure}

In total, 100 random keys (\textit{b}) were created for each group/key size. Each secret key is profiled only once and the success rate of each single-trace key recovery attack is recorded. Interestingly, the time difference between consecutive square operations and square+multiply instances increases when the group size becomes larger. It means that higher group sizes give better resolution for the attacker. For instance, there is a 20,000-25,000 cycle difference between each square when the group size is set to 6,144, which is much larger than the size of 1,024, leading to longer traces for an attacker to extract the key. Hence, the attacker adjusts the key extraction process based on the time difference distribution between the activities. There are seven patterns that need to be distinguished based on the time differences: "0", "1", "11", "1X1", "1XX1", "1XXX1", "1XXXX1". The time difference increases for each pattern as more operations are involved, leading to more samples without any activity between each cache miss as shown in Figure~\ref{fig:SRP}.

\begin{table}[h]
\small
\centering
\caption{Leakage rates obtained with Mastik~\cite{yarom2016mastik} and Prime+IStore attacks for different group sizes in the SRP\_Calc\_Server\_Key function for the OpenSSL-1.1.1w implementation.}
\setlength{\tabcolsep}{5pt}
\begin{tabular}{l|c|c|c|c|}
\cline{2-5}
& \multicolumn{4}{c|}{\textbf{Group Size (bits)}} \\
\cline{2-5}
& \textbf{1024} & \textbf{2048} & \textbf{4096} & \textbf{6144} \\\hline\hline
\multicolumn{1}{|l|}{\textbf{Prime+IStore}} & 65\% & 73\% & 83\% & 90\% \\\hline
\multicolumn{1}{|l|}{\textbf{Mastik (PnP)}} & 22\% & 31\% &  37\% & 48\%  \\\hline
\end{tabular}
\label{tab:SRP_leak}
\end{table}

An attacker can leak up to 90\% of the possible bits in a given secret exponent with the Prime+IStore attack, which is calculated over 100 keys. The highest leakage rates are achieved on the largest group size as the time difference between each pattern increases significantly, leading to less noisy traces. The lowest leakage rate is 65\% for the key size of 1024 as given in Table~\ref{tab:SRP_leak}. We also compare the Prime+IStore attack with the Prime+Probe (PnP) attack from the Mastik tool. The PnP attack is noisier compared to our attack because L1i cache hits and misses cannot be reliably distinguished, as shown in Figure~\ref{fig:CPUcycleTime_cascade}. Hence, we selected a threshold for the miss/hit classification by matching the expected number of cache misses to the actual cache misses obtained from reverse engineering. Moreover, the number of outliers is much higher compared to the Prime+IStore attack. As a result, PnP can leak up to 48\% of the possible bits, which is considerably lower than our attack. Our results show that with the Prime+IStore technique a single trace attack can be performed with high accuracy.

\begin{table*}[t]
\centering
\caption{Comparative analysis of SMC-ISpectre Attack Vulnerability across diverse Microarchitectures through various strategies. An $\times$ indicates unsupported instructions. Secrets leaked without SMC behavior are marked with $\LEFTcircle$, while successful SMC-ISpectre attacks are denoted by $\CIRCLE$.}
\setlength{\tabcolsep}{7pt}
\scalebox{0.65}{
\begin{tabular}{c|c|c|c|c|c|c|c|c|c|c}
\textbf{}            & \textbf{Westmere EP} & \textbf{Sandy Bridge} & \textbf{Ivy Bridge} &\textbf{Broadwell}  & \textbf{Ice Lake} &\textbf{Cascade Lake}&\textbf{Comet Lake}& \textbf{AMD Ryzen 5} &\textbf{AMD EPYC 7232P}&\textbf{Tiger Lake}\\ \hline
\textbf{Load}        & $\LEFTcircle$ & $\LEFTcircle$ & $\LEFTcircle$ & $\LEFTcircle$ & $\LEFTcircle$ & $\LEFTcircle$& $\LEFTcircle$ & $\LEFTcircle$ & $\LEFTcircle$& $\LEFTcircle$\\ \hline
\textbf{Flush}       & $\CIRCLE$ & $\CIRCLE$ & $\CIRCLE$ & $\CIRCLE$ &  $\CIRCLE$ & $\CIRCLE$ & $\CIRCLE$ & $\CIRCLE$ &$\LEFTcircle$& $\CIRCLE$\\ \hline
\textbf{FlushOPT}    & $\CIRCLE$ & $\CIRCLE$ & $\times$ & $\times$  & $\CIRCLE$ & $\CIRCLE$ & $\CIRCLE$ & $\CIRCLE$ & $\LEFTcircle$& $\CIRCLE$\\ \hline
\textbf{Store}       & $\CIRCLE$ & $\CIRCLE$ & $\CIRCLE$ & $\CIRCLE$  & $\CIRCLE$ & $\CIRCLE$ & $\CIRCLE$ & $\CIRCLE$& $\CIRCLE$& $\CIRCLE$\\ \hline
\textbf{Lock}     & $\CIRCLE$ & $\CIRCLE$ & $\CIRCLE$ & $\CIRCLE$  & $\CIRCLE$ & $\CIRCLE$ & $\CIRCLE$ & $\CIRCLE$& $\CIRCLE$& $\CIRCLE$\\ \hline
\textbf{Prefetch}    & $\Circle$ & $\Circle$ & $\Circle$ & $\CIRCLE$  & $\Circle$ & $\CIRCLE$ & $\CIRCLE$ &$\LEFTcircle$& $\LEFTcircle$& $\Circle$ \\ \hline
\textbf{PrefetchNTA} & $\Circle$ & $\Circle$ & $\Circle$ & $\Circle$  & $\Circle$ & $\LEFTcircle$ & $\LEFTcircle$ &$\LEFTcircle$& $\LEFTcircle$ & $\Circle$ \\ \hline
\textbf{Execute}     & $\Circle$ & $\Circle$ & $\Circle$ & $\Circle$  & $\Circle$ & $\Circle$ & $\Circle$ & $\Circle$ & $\Circle$& $\Circle$\\ \hline
\textbf{Clwb}     & $\times$ & $\times$ & $\times$ & $\times$  &$\times$ & $\CIRCLE$ & $\Circle$ & $\Circle$ & $\LEFTcircle$ & $\CIRCLE$ \\ \hline
\end{tabular}
}
\label{table:Spectre_attacks}
\end{table*}

\subsection{Case Study IV: ISpectre Attack}\label{subsec:case4}
In this section, we leverage Flush+iReload attacks in the context of the Spectre v1. 
We exploit the Pattern History Table (PHT)~\cite{canella2019systematic}, which predicts the outcome of conditional branch instructions.
The original Spectre v1 variant is designed to leak information from the data cache by accessing out-of-bound array elements and encoding the secret value to different memory pages.
However, \texttt{ISpectre} attack exploits an indirect \texttt{call} instruction based on the value of the shared oracle code page array, which not only aims to access unauthorized memory but also aims to execute the code located at an address that is determined speculatively by the branch prediction unit as given in Listing~\ref{lst:conditional_assembly}.
Since we leak the secret from the L1i cache sets, traditional shared cache-focused defenses are insufficient~\cite{ren2021see} against ISpectre.

\noindent\textbf{Experiment Setup.} \texttt{ISpectre} experiments are conducted on various microarchitectures as shown in Table~\ref{table:Spectre_attacks}. In total, we test 8 Intel and 2 AMD microarchitectures to evaluate our attack. All experiments are conducted on the Ubuntu 20.04.6 LTS operating system with default configuration. All microarchitectures (microcode code 0x5003707 in Intel Cascade Lake) are patched with their vendors' latest Spectre defense mechanisms.

\noindent\textbf{ISpectre.} In the \texttt{ISpectre} attack, the branch instruction in the \texttt{victim\_function} is mistrained with in-bound values and the indirect function call is executed by accessing \texttt{oracle\_code\_page} with the predefined offset size (notsecret[index] $\times$ cache line size).  
Next, the adversary supplies an out-of-bound \texttt{index} value, leading to the execution of an out-of-bound memory location with the predefined offset.
Once the victim function executes the instruction in this memory region, the cache line is speculatively brought to the L1i cache.
This behavior acts as the encoding of the secret into an instruction cache line.
Next, the attacker performs the Flush+iReload profiling on all the instruction cache lines in the \texttt{oracle\_code\_page} and measures the execution time.
If the execution time is higher than a threshold, the secret character is already encoded into that offset in the \texttt{oracle\_code\_page}, which was brought to the L1i cache.

\begin{lstlisting}[language=C, caption={C Code snippet for the victim function executing an indirect branch instruction}, label=lst:conditional_assembly,basicstyle=\ttfamily\scriptsize, basewidth={.48em}, backgroundcolor=\color{white}]
asm volatile( "call *%0\n" 
: : "c"(oracle_code_page + notsecret[index] * CACHE_LINE_SIZE):"rax");}
\end{lstlisting}

\noindent\textbf{Evaluation.} 
Our attack can exploit \texttt{flush}, \texttt{flushopt}, \texttt{store}, \texttt{lock}, and \texttt{prefetch} instructions to create SMC conflict and distinguish the execution time between L1i cache hit and DRAM access on various x86 microarchitectures as given in Table~\ref{table:Spectre_attacks}. 
The attacker measures the time it takes to execute each cache line associated with the secret byte and distinguishes the correct secret.
Even though \texttt{load}, \texttt{prefetchnta}, and \texttt{execute} instructions do not create an SMC conflict, \texttt{load} can still leak secret characters, which is illustrated with a half circle in Table~\ref{table:Spectre_attacks}. 
Both \texttt{lock} and \texttt{store} instructions are successful at leaking the secret bytes in all x86 microarchitectures. 
As the \texttt{clwb} instruction is not available in older generations of Intel microarchitectures, it is not suitable for the Spectre attacks. 
The \texttt{flush} operations are also successful at leaking the secret bytes except for the AMD EPYC 7232P microarchitecture, in which the flush instructions have no effect on the SMC conflicts.

The highest leakage rate is 3161 B/s with the \texttt{lock} instruction as given in Table~\ref{tab:spec_leak}. We also achieve up to 3109 B/s leakage rate with the Flush+iFlushopt attack with an average success rate of 98.37\%. Furthermore, the \texttt{Flush+iFlushopt} attack reveals secret bytes with 4105.84 B/s on AMD Ryzen 5. This performance surpasses TLB-Evict+Prefetch attack~\cite{lipp2022amd}, which recovers 96.7\% rate at 58.98 B/s, and outperforms Lipp et al.~\cite{lipp2020take} of 0.66B/s on the same AMD microarchitecture.
Notably, non-SMC Spectre attacks are also feasible to leak secret bytes in the L1i cache. 

\begin{table}[h]
\small
\centering
\caption{Spectre-v1 leakage rates for Flush+iReload attacks. The leakage rates are given in B/s. The experiments are conducted on Intel Cascade Lake and AMD Ryzen 5 processors.}
\setlength{\tabcolsep}{5pt}
\scalebox{0.9}{
\begin{tabular}{l|c|c|c|c|c|c}
\textbf{Processor} & \textbf{Flush} & \textbf{Flushopt} & \textbf{Store} & \textbf{Lock} & \textbf{Prefetch} & \textbf{Clwb} \\\hline
\textbf{Intel CL} & 1513 & 3109 & 2039 & 3161 & 1556 & 
 2974 \\\hline
\textbf{AMD Ryzen} & 4045 & 4105 & 1382 & 2168 & N/A & N/A \\\hline
\end{tabular}
}
\label{tab:spec_leak}

\end{table}

\section{Countermeasures}\label{sec:countermeasures}

\subsection{Dynamic Detection}\label{sec:dynamic_detection}

Both Intel and AMD processors implement a diverse set of hardware performance counters to monitor running processes~\cite{zhang2016cloudradar,briongos2018cacheshield,gulmezoglu2019fortuneteller}. We create a dynamic detection tool by profiling diverse benign applications from the Phoronix-benchmarks suite~\cite{OpenBenchmarkingTests}. 
In the profiled counter list, we include counters used to detect cache attacks and Spectre attacks~\cite{zhang2016cloudradar,briongos2018cacheshield,gulmezoglu2019fortuneteller} as well as our proposed counters related to SMC conflicts mentioned in Section~\ref{sec:SMCAttack}. 
We observed that performance counters used in previous studies are insufficient to detect SMC-based attacks. Specifically, Prime+iProbe attacks are less impacted by mispredicted branch instruction counters than the Spectre attacks, which directly induce branch mispredictions. Consequently, the F-score of the detection model for the Prime+iProbe attacks using \texttt{BR\_\allowbreak MISP\_\allowbreak RETIRED.ALL\_\allowbreak BRANCHES} is 0.8000, which is lower than 0.9215 achieved for Flush+iReload attacks. Furthermore, the LLC miss counter event~\cite{gulmezoglu2019fortuneteller} faces a challenge in detecting our \texttt{Prime+iProbe} activity since our attack evicts cache lines from the L1i cache and has minimal effect in LLC, resulting in a 0.7679 F1 score. Hence, we selected counters more related to SMC conflicts on Intel architectures.

We collected 20 benign benchmarks and 12 attack executions on the Intel Cascade Lake microarchitecture. For the malicious dataset, we profile \texttt{Prime+iProbe} and \texttt{Flush+iReload} variants, triggering SMC conflicts (Table~\ref{table:Spectre_attacks}). The datasets are collected for a duration of 10 seconds with 100 ms resolution, generating 100 measurements for both execution types. 
The train and test datasets are split with 80\% and 20\% ratios for the evaluation. We identified that \texttt{cycle\_\allowbreak activity.\allowbreak stalls\_\allowbreak total}, \texttt{machine\_clears.count}, and \texttt{machine\_\allowbreak clears.\allowbreak smc} counters yield high F1 scores. 6 different \texttt{Prime+iProbe} variants are detected with 99.36\% accuracy and F-score of 0.9870 and 0.85\% FPR by employing the \texttt{machine\_clears.smc} counter.
The reason behind it's not 100\% accurate is that some of the measurements from the \texttt{amg} benchmark are treated as malicious applications since \texttt{amg} benchmark creates significant SMC behavior, unlike other benign applications. Additionally, \texttt{Flush+iReload} attacks are detected with 100\% accuracy by utilizing \texttt{machine\_clears.smc} counter. Our results demonstrate that employing an SMC-related detection tool can successfully distinguish \texttt{Prime+iProbe} and \texttt{ISpectre} attacks. 

\subsection{Software and Hardware Countermeasures}
The vulnerabilities described in this study can be avoided by transforming the vulnerable implementation to a \emph{constant-time} version, i.e., without secret-dependent branches or memory accesses.
\texttt{Raccoon}~\cite{rane2015raccoon}, \texttt{Escort}~\cite{rane2016secure}, and \texttt{EncLang}~\cite{sinha2017compiler} offer methods to automatically transform existing code into a constant-time representation. However, they require source code annotation and may significantly impact performance, e.g., when ORAMs are employed. Additionally, there has been significant research in implementing automated verification of constant-time properties. Tools like \texttt{ct-verif}~\cite{almeida2016verifying} and \texttt{CacheD} use static code analysis or symbolic execution to prove the absence of side-channel vulnerabilities. Since this does not scale well with increasing implementation sizes, various dynamic approaches have been proposed, e.g., \texttt{ctgrind}~\cite{langley2010ctgrind}, which simply keeps secret memory uninitialized. \texttt{DATA}~\cite{weiser2018data} and \texttt{MicroWalk}~\cite{wichelmann2018microwalk} evaluate implementations by comparing whole execution traces and quantifying leakage while pinpointing leaking instructions. 
A straightforward but costly countermeasure is disabling SMT altogether. This way, core resources like pipeline and L1 caches are not shared, which are exploited by the described attacks and many others~\cite{aciiccmez2007yet,aciiccmez2010new}. However, this will also lead to significant performance degradation. 

\section{Discussion and Limitations}\label{sec:discussion}

\noindent\textbf{SMC-creating instructions.} We profiled potential instructions that can invalidate the instruction cache lines available in x86 ISA, which may lead to SMC conflicts in both Intel and AMD devices. However, there might be more instructions than the profiled instructions in this work. These instructions can be discovered with a sophisticated fuzzing framework by either observing the timing values or performance counters~\cite{weber2021osiris}. It would also be interesting to analyze other machine clear events described in Ragab et al.~\cite{ragab2021rage} to determine their effect on timing measurements.

\noindent\textbf{Performance counter-based detection.} Performance counter-based detection systems are vulnerable to evasive attacks as sophisticated attackers can modify their attack code behavior to bypass detection systems~\cite{kosasihsok}. We believe that SMC-based attacks will always increment the machine clears event compared to benign applications, which is more difficult to hide compared to cache misses and stall cycle counters.

\noindent\textbf{Comparison with Previous Instruction Cache Attacks.} There are several attacks targeting the instruction cache~\cite{aciiccmez2007yet,aciiccmez2010new,yarom2016mastik} and $\mu$op cache~\cite{kim2021uc,ren2021see}. These attacks achieve higher bandwidth than the proposed SMC attacks when they are used for covert channels since accessing instructions in the L1i cache takes more time than for the SMC attacks. However, there are two advantages of the proposed SMC attacks: 1) The sibling virtual core slows down in parallel, which increases the time difference between secret activities, as shown in the RSA and SRP attack case studies. The slow-down on the sibling virtual core reaches up to 10 times, compensating for the time spent on the L1i cache accesses compared to previous work. 2) The uncertainty amount between an L1i cache hit and miss in the SMC attacks is significantly less than other attacks since they can only achieve a 1-2 cycle difference, missing important secret-specific actions. Moreover, some environments, such as AMD processors, do not provide high-resolution timers. For instance, the rdtsc instruction has only 20 cycles resolution, which cannot be used to distinguish L1i cache hit and miss for the Mastik tool. Hence, the single-trace attack with Prime+IStore was able to achieve a higher success rate compared to the Prime+Probe attack with the Mastik tool.
\section{Related work}\label{sec:relatedwork}

\noindent\textbf{SMC-based Attacks.} There are two previous works on the usage of SMC conflicts. The first work~\cite{aldaya2022hyperdegrade} utilizes the clflush instruction on the shared libraries by repeatedly flushing the executed instructions. The authors slow down the victim process running in the sibling core and increase the resolution of the Flush+Reload attack. This work shows that an attacker can slow down the sibling core 43 times due to a high number of machine clears and cache misses. On the other hand, our attack shows that an attacker can leak secret keys without shared libraries by creating eviction sets in the L1i cache. The second work~\cite{ragab2021rage} focuses on extending the speculative window size by creating several machine clear events including SMC. The secret is leaked through the Flush+Reload cache covert channel through the data cache, while our attack leaks the secrets from the L1i cache covert channel from an indirect branch.

\noindent\textbf{Instruction Cache Attacks.}
Aciicmez et al.~\cite{aciiccmez2007yet} showed that Prime+Probe attacks could be implemented on the L1 instruction cache to monitor the executed instructions from the victim process in the sibling core. 
Moreover, Aciicmez et al.~\cite{aciiccmez2010new} build upon earlier work~\cite{aciiccmez2007yet} in instruction cache attacks, providing more effective ways to explore the L1I cache incorporates Vector Quantization (VQ) and Hidden Markov Models (HMM) techniques. This work recovered the DSA private key utilizing lattice methods and proposed a mitigation method by disabling multi-threading, disabling caching, cache flushing, and arranging the memory layout with dummy \texttt{nop} instructions.

\noindent\textbf{Spectre Attacks.} A large body of Spectre attacks has been explored in the security community~\cite{xiong2021survey,canella2019systematic}. These either attacks focus on manipulating branch prediction mechanisms~\cite{kocher2020spectre,chen2019sgxpectre,koruyeh2018spectre,van2020lvi} or establishing new covert channels~\cite{bhattacharyya2019smotherspectre,loughlin2021dolma,kocher2020spectre}. 
Smotherspectre~\cite{bhattacharyya2019smotherspectre} creates a side channel to leak secrets based on the executed instructions. The covert channel makes use of the execution ports assigned to distinct instruction types. DOLMA~\cite{loughlin2021dolma} shows that data TLB structure can be leveraged to leak secrets in the transient domain if the secret is encoded into a separate page. The first examples of Spectre attacks~\cite{kocher2020spectre} leverage the data cache to leak secrets through data caches.   
\section{Conclusion}\label{sec:conclusion}
In this paper, we present new Prime+Probe and Flush+Reload attacks on the instruction cache created by the SMC detection mechanism in x86 processors. We show that an attacker can implement high-resolution attacks to leak RSA decryption keys with only 10 measurements. Moreover, single-trace attacks are possible with SMC-based attacks on an OpenSSL library. A new version of the Spectre attack is proposed to leak secrets in the memory. Finally, we propose a hardware performance counter-based detection system to detect ongoing SMC-based attacks with an F-score of 0.98 with a minimal performance overhead. The artifacts supporting our experiments are publicly available on our GitHub repository\footnote{\url{https://github.com/hunie-son/SMaCk}}.
\begin{acks}
We thank our shepherd, Adam Morrison, and anonymous reviewers of ASPLOS 2025 for their valuable feedback.
SMaCk is reported to Intel and AMD security incident teams, and AMD published a security bulletin (AMD-SB-7024). We also reported the security bug in the OpenSSL library to the OpenSSL community. 
\end{acks}
\bibliographystyle{ACM-Reference-Format}
\balance
\bibliography{Ref}
\newpage

\end{document}